\documentclass[preprint]{aastex}
\usepackage{amsmath}
\newcommand{\beq}{\begin{equation}}
\newcommand{\eeq}{\end{equation}}
\newcommand{\bea}{\begin{eqnarray}}
\newcommand{\eea}{\end{eqnarray}}
\newcommand{\rem}[1]{ }

\begin{document}
\title{Modeling Spectral Variability of Prompt GRB Emission within the Jitter Radiation Paradigm}

\author{Mikhail V. Medvedev$,^{1,2}$ Sriharsha S. Pothapragada$,^1$ Sarah J. Reynolds$^1$} 

\affil{$^1$ Department of Physics and Astronomy, University of Kansas, Lawrence, KS 66045}

 \affil{$^2$ Institute for Nuclear Fusion, RRC ``Kurchatov Institute", Moscow 123182, Russia}

\begin{abstract}
The origin of rapid spectral variability and certain spectral correlations of the prompt gamma-ray burst emission remains an intriguing question. The recently proposed theoretical model of the prompt emission is build upon unique spectral properties of jitter radiation --- the radiation from small-scale magnetic fields generated at a site of strong energy release (e.g., a relativistic collisionless shock in baryonic or pair-dominated ejecta, or a reconnection site in a magnetically-dominated outflow). Here we present the results of implementation of the model. We show that anisotropy of the jitter radiation pattern and relativistic shell kinematics altogether produce effects commonly observed in time-resolved spectra of the prompt emission, e.g., the softening of the spectrum below the peak energy within individual pulses in the prompt light-curve, the so-called ``tracking" behavior (correlation of the observed flux with other spectral parameters), the emergence of hard, synchrotron-violating spectra at the beginning of individual spikes. Several observational predictions of the model are discussed.
\end{abstract}

\section{Introduction}

\begin {figure}
\includegraphics[angle = 0, width = 0.95\columnwidth]{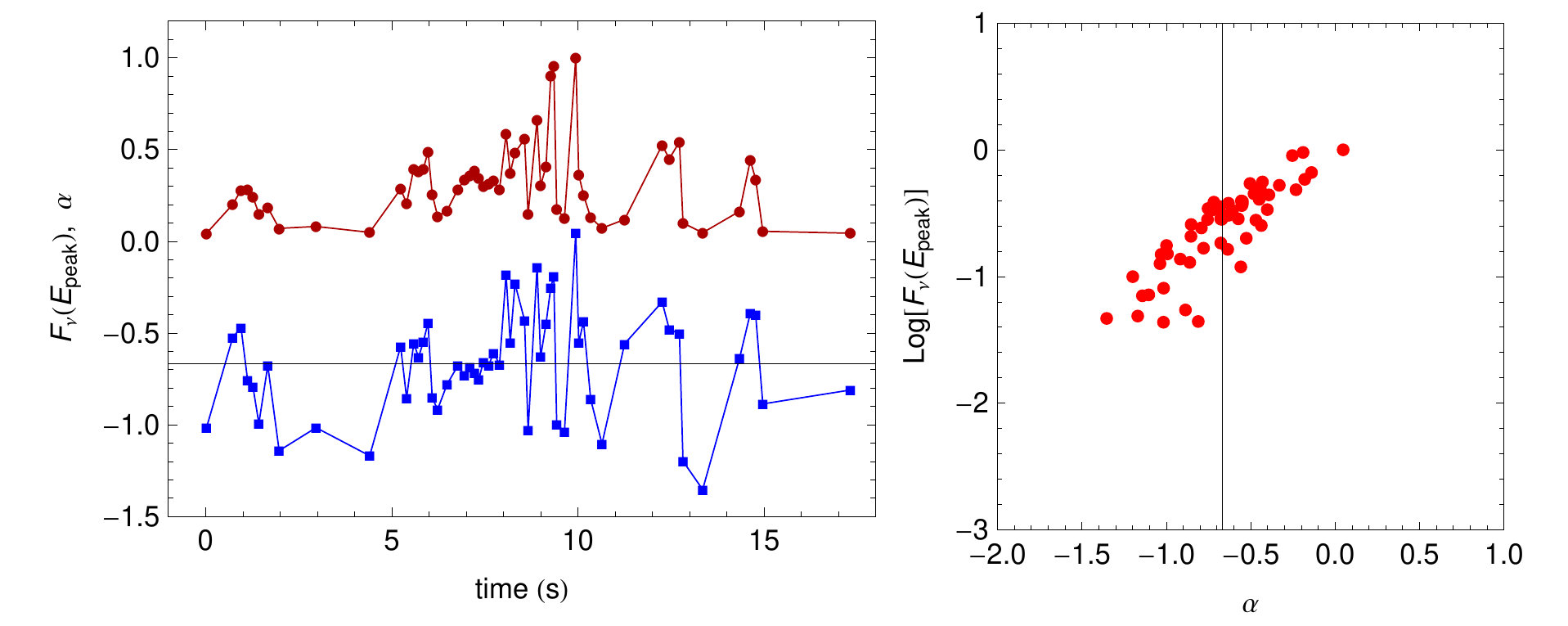}
%\plotone{trigger6124.pdf}
\caption{ The tracking behavior in BATSE burst 6124. Note that when $F_{\nu}$ at the peak energy  (top, dark red curve) is high, the photon index $\alpha$ (bottom, blue curve) exceeds the synchrotron limit $\alpha\le-2/3$ shown with the vertical line in the left panel.}
\label{trigger}
\end {figure}

Gamma-ray burst (GRB) prompt emission exhibits rapid spectral variability the origin of which is not fully understood yet. While some of it --- the hard-to-soft evolution at late prompt stages --- can be attributed to the high-latitude emission in the standard synchrotron shock model \citep{KP00,Granot+09}, the long-acknowledged ``tracking'' behavior \citep{Bhat+94,Crider+97,Frontera+00} remains unexplained within this model. The tracking behavior is quite intriguing: the low-energy spectral index $\alpha$ follows the observed flux \citep{Crider+97,Harsha05}, with harder index corresponding to higher flux, as is illustrated in figure \ref{trigger} (we used data from \citealp{Kaneko+06}). To our knowledge, the synchrotron shock model (e.g., with self-absorption and comptonization) does not seem to generally predict such a correlation. The situation becomes even more dramatic when one takes into account that the hardest observed $\alpha$'s cannot be reconciled with the standard synchrotron theory, which constrains the lower energy photon index to be $\alpha\le-2/3$ \citep{Preece+98,Preece+00}, and that the most of the GRB spectra exhibit $\alpha\sim-1$ --- the value by no means special within the theory \citep{Kaneko+06}.\footnote{It is worth mentioning that BATSE time-resolved spectral catalog \citep{Kaneko+06} is the largest available to date and will remain such for at least a few years. Besides, GBM instrument onboard of Fermi-GRO is very similar to BATSE in many respects, hence it is natural to compare the results with BATSE data.}

A recent theoretical model \citep{M06} of the spectral variability of prompt GRBs predicts that the above properties naturally follow from the jitter radiation theory \citep{M00} of electron emission in small-scale magnetic fields generated {\it in situ}. Note that one should not use here the results by \citep{Fleish06}, as we discuss at the end of Section \ref{s:model}. Here we do not limit ourselves to a particular model of a GRB: either a kinetically-driven outflow, including baryon-dominated and lepton-dominated ones (i.e., with electron-ion and electron-positron pair plasmas, respectively), or a magnetic-field-dominated outflow (e.g., driven by Poynting flux), or any combination thereof. 

The prediction that relativistic shocks generate strong magnetic fields via the Weibel-like (particle streaming) instability \citep{ML99} has been confirmed with a large number of numerical PIC simulations in both baryonic and pair plasmas  \citep{Silva+03,Fred+04,Nish+05,Spitk08,CSA08,KKSW08}. The generated fields are random on a very small (sub-Larmor) scale, hence the synchrotron theory is inapplicable in such environments while the jitter radiation seems to be validated in these simulations \citep{Hededal05}, whose immediate relevance to GRB physics has recently been strengthened \citep{MS09}. The generation of small-scale magnetic fields has also been discovered in PIC simulations of magnetic reconnection in pair plasmas \citep{Swisdak+08,ZH08}. We can speculate that a strong energy release, as in GRBs, can produce electron-positron plasmas {\it in situ} during a reconnection event, even if the plasma is initially lepton-poor. The strong small-scale fields are produced via the Weibel instability by the streams of accelerated particles in the reconnection exhaust funnels, hence jitter radiation is expected to be produced here. 

In this paper, we study spectral variability and correlations within the jitter radiation framework using a simple numerical model of an instantaneously illuminated, relativistically expanding thin spherical shell segment within a jet \citep{Harsha+07,Sarah+08}. We interpret the flux--$\alpha$ correlation as a combined effect of temporal variation of the viewing angle, the anisotropy of the radiation field in the shell and relativistic aberration. We mostly focus on the $F_{\nu}(E_{\rm peak})-\alpha$ correlation because other parameters --- the hard index, the peak energy itself $E_{\rm peak}$, and the bolometric flux --- are more susceptible to the overall GRB energetics, details of electron acceleration, etc.; so, unlike $\alpha$, more factors can affect them and cause mutual correlations.

\section{Numerical model}
\label{s:model}

\begin {figure}
\plotone{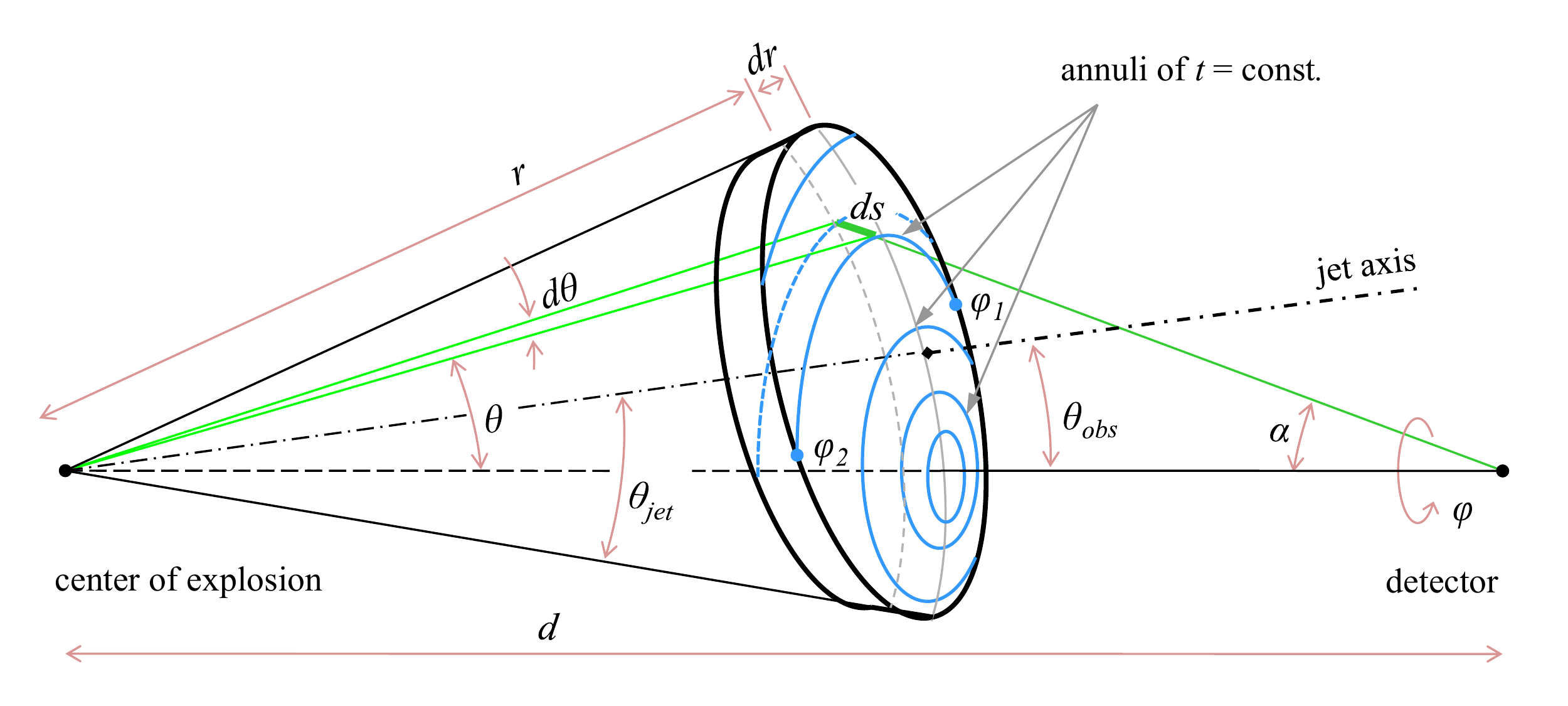}
\caption{Cartoon showing the radiating shell surface of the mis-aligned jet. Photons emitted from thin annuli or parts thereof centered on the line of sight arrive at a detector simultaneously.}
\label{jetcone} 
\end {figure}

\begin{figure}
 \centering
\plotone{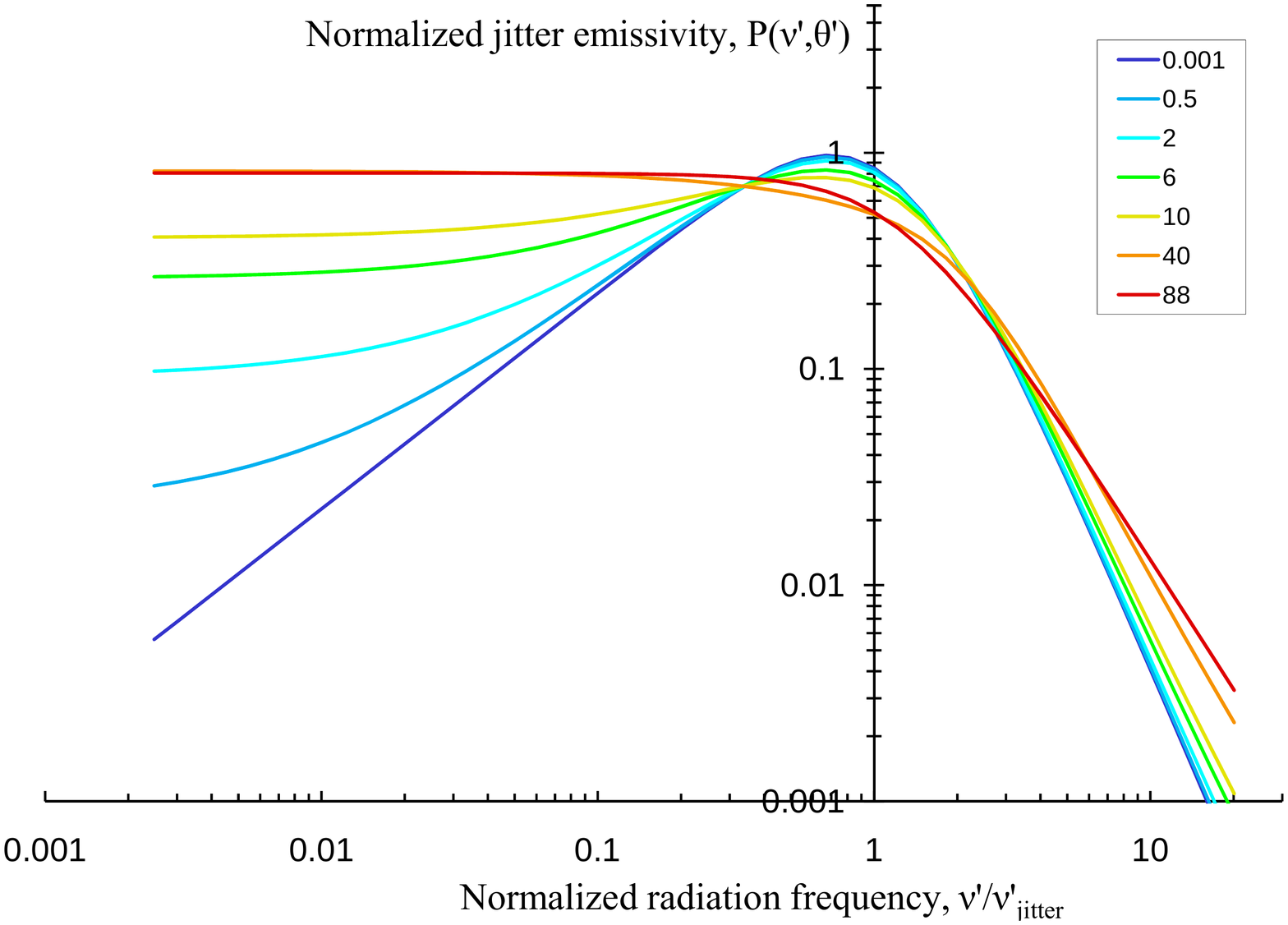}
 \caption{Co-moving normalized jitter radiation emissivity of an electron averaged over the electron ensemble, $P(\nu',\theta')$, at angles (from hard to soft, blue through red) $\theta'=0.001,\ 0.5,\ 2,\ 6,\ 10,\ 40,\ 88$~degrees. The frequency is normalized to the jitter peak frequency, $\nu'_{\rm jitter}=\gamma_e^2c/\lambda_B$.}
 \label{srcspec}
\end{figure}

One usually assumes that the each individual ``spike'' or ``pulse'' in the prompt light-curve is a single ``emission episode" produced by an internal dissipation event (a shock or reconnection) that occurs within the ejecta when two thin plasma shells of different speeds and/or field orientations collide. The emission region may generally be curved, hence we use the model of the instantaneously illuminated thin spherically expanding relativistic shell. Because of the shell curvature, the photons emitted along the line of sight, $\theta'=\theta=0$, see Figure \ref{jetcone}, arrive at the detector first whereas the high-latitude emission, $\theta'\ga1$, comes later. [Hereafter, `primed' variables are measured in the shell co-moving (i.e., source) frame and `unprimed' ones are in the observer's (detector) frame]. Hence, an instantaneous `flash' in the source frame translates to a light curve in the observer's frame (we neglect the cosmological redshift effect, assuming $z=0$). The angle-dependent spectrum (i.e., the volume emissivity, $j'_{\nu'}=n_e' P(\nu',\theta')$, $n_e'$ being the electron density and $P$ being the ensemble-averaged spectral power per electron) of jitter radiation in the source frame, shown in Figure \ref{srcspec}, thus translates into a time-dependent spectrum in the detector frame. Integration over the emitting volume and accounting for time dilation, relativistic Doppler effect and relativistic angle aberration yields the energy flux $F_\nu$ seen by a distant observer. 

Let us consider emission from a thin segment of a spherical shell expanding radially with the Lorentz factor $\Gamma=(1-\beta^2)^{-1/2}$, $\beta=v/c$ and confined within a jet of opening angle $\theta_{\rm jet}$, its axis can be misaligned with the line of sight through an angle $\theta_{\rm obs}$, as shown in Figure \ref{jetcone}. The observed flux from an optically thin relativistic source has been calculated by \citet{WL99, GPS99, RP02} and others. Here we briefly outline the method following \citet{WL99}. We use spherical coordinates with a GRB explosion being at the origin; the line-of-sight (horizontal axis, $\theta=0$) is from the origin toward the detector located some large distance $d\gg r$ away from it. If the detector plane is orthogonal to the line-of-sight and a ray makes an angle $\alpha$ with it, the observed energy flux through a unit surface per frequency is
\beq
F_\nu=\int I_\nu(\alpha,\phi)\cos\alpha d\Omega
\approx \int d\phi\int I_\nu(\alpha,\phi) \alpha d\alpha, 
\eeq
where $\cos\alpha$ comes from the obliqueness of the ray to the detector plane, which is negligible at small angles: $\cos\alpha\approx 1$; also $\sin\alpha\approx\alpha$. Here $I_\nu(\alpha,\phi)$ is the intensity along the ray in the direction $(\alpha,\phi)$ and $d\Omega$ is the differential solid angle of the emitting region. From simple geometry 
$\alpha\approx r\sin\theta/d=(r/d)(1-\mu^2)^{1/2}$, where $\mu=\cos\theta$, so that for given $r$,
$\alpha d\alpha=(-r^2/d^2)\mu d\mu.$

The intensity is calculated as follows,
\beq
dI_\nu=j_\nu ds=j'_{\nu'}{\cal D}^2(\mu)dr/\mu,
\eeq
where the differential path length is $ds=(r^2d\theta^2+dr^2)^{1/2}=dr/\mu$ and
we used the Lorentz invariance of the quantity $j_\nu/\nu^2$, so that $j_\nu/\nu^2=j'_{\nu'}/\nu'^2$. The observed and emitted frequencies, $\nu$ and $\nu'$, are related as
\beq
\nu=\nu'{\cal D}(\mu)=\frac{\nu'}{\Gamma(1-\beta\mu)}
\eeq 
with ${\cal D}=[\Gamma(1-\beta\mu)]^{-1}=\Gamma(1+\beta\mu')$ being the relativistic Doppler boosting factor. The emissivity in the co-moving frame, $j'_{\nu'}({\bf \Omega}'_d,{\bf r},t_e)$, measured in erg/(s~cm$^2$~Hz~sr), is a function of frequency $\nu'$, position ${\bf r}=(r,\theta,\phi)$, time $t_e$, and the direction toward the detector measured in the co-moving frame ${\bf \Omega}'_d=(\theta',\phi')$, where the angles transform due to relativistic aberration as 
\beq
\phi'=\phi, \quad \cos\theta'=\mu'=\frac{\mu-\beta}{1-\beta\mu}.
\eeq 
Because of the shell curvature, the photon arrival time, $t$, is related to the emission time, $t_e$, as $t=t_e-r\mu/c$, so that $t=0$ is for a photon emitted from the origin at $t_e=0$. Combining the above equations together, we arrive at a general result analogous to equation (8) of \citep{WL99},
\beq
F_\nu(t)=\frac{1}{d^2}\int d\phi\int d\mu \int r^2dr\, {\cal D}^2 j'_{\nu/{\cal D}}({\bf \Omega}'_d,{\bf r},t+r\mu/c),
\label{Fnu-gen}
\eeq
where limits of integration are set by the geometry of the source, $(\phi,\mu)\in{\rm jet}$. In general, $\Gamma$ is a function of position and time. Our numerical model calculates the spectral flux for an arbitrarily shaped source and $\Gamma(\theta',\phi')$. For a jet, one gets $\theta\in[0,\theta_{\rm jet}+\theta_{\rm obs}]$, $\phi\in[\phi_1(\theta),\phi_2(\theta)]$, see Figure \ref{jetcone}. The $\phi$-limits are found numerically for each $\theta$ by ensuring  $\phi$ to be within the edges of jet cone; in the absence of edges $\phi\in[0,2\pi]$. For simplicity, we also assume $\Gamma={\rm const}$ hereafter. Assuming that emission occurs at $r=R$ and $t_0=t_e=R/\beta c$ and the shell is of infinitesimal thickness, $\Delta R\to0$, we take the emissivity as $j'_{\nu'}=n_e' P(\nu',\theta')\Theta(\Delta R'/2-|r'-R'|)\delta(t'_e-t'_0)\to \Sigma'_e P(\nu',\theta')\delta(r-R)\delta(t_e-t_0)$, where $\Theta$ and $\delta$ are the Heaviside step-function and Dirac $\delta$-function. Here we used that  $r'=\Gamma r$, $t'=t/\Gamma$ and the identities $\delta(ax)=|a|^{-1}\delta(x)$, $\Theta(ax)={\rm sign}(a)\Theta(x)$, and also introduced the electron surface density $\Sigma'_e=n_e'\Delta R'$. Substituting $j'_{\nu'}$ into eq. (\ref{Fnu-gen}) and integrating over $dr$ one obtains the time-dependent $\delta$-function $\propto\delta(\mu-\beta^{-1}+tc/R)$. Upon integration over $d\mu$ and $d\phi$ we obtain 
\beq
F_{\nu}(t)=F_0\,{\cal D}^2(t)\, P\!\left(\nu'(t),\mu'(t)\right)\left(\phi_2(t)-\phi_1(t)\right),
\eeq
where $F_0,$ is the normalization constant: $F_{\nu}(t_e)=1$ at the peak frequency, ${\cal D}(t)=R/(\Gamma\beta ct)$, $\nu'(t)=\nu/{\cal D}(t)$, $\mu'(t)=({\cal D}(t)\Gamma^{-1}-1)/\beta$, and $\phi_1, \phi_2$ are found numerically, as is described above. For a conical jet, $\phi_2-\phi_1$ have been calculated analytically \citep{IN01,YIN02}.

The jitter comoving spectrum $P(\nu',\theta')$ depends on the full three-dimensional spatial spectrum of the magnetic field and the electron distribution in the emitting shell. It is calculated elsewhere in full detail, see \citet{M00,M06,Sarah+08,Sarah+09}. In order to keep our model simple, we assume the field spatial spectra in the shell (e.g., shock plane) and normal to it (i.e., along the Weibel current filaments) to be identical and described by a broken power-law peaked at $k_B\sim\lambda_B^{-1}$ --- the characteristic field correlation scale. The electron energy distribution\footnote{The electron distribution can be anisotropic, so more emission can be produced at small $\theta'$. At present, no good {\em quantitative} model of the electron angular distribution is available, so $n\propto\cos(\theta')$ can be a good alternative to the isotropic model.} is assumed to be a power-law with the low energy cutoff at $\gamma_e$, $n(\gamma)\propto \gamma^{-p}$ for $\gamma\ge\gamma_e$. Hence the peak jitter frequency is $\nu'_{\rm jitter}=\gamma_e^2c/\lambda_B$. The resultant spectrum is shown in figure \ref{srcspec} for several angles $\theta'$ ranging from 0 (head-on shell/shock) to $\pi/2$ (edge-on shell/shock in the comoving frame), at a higher angle $\theta'>\pi/2$ the spectrum is identical to that at the angle $\pi-\theta'$. The $P(\nu',\theta')$ spectrum above the peak is set either by the electrons, yielding the slope $-(p-1)/2$, or by the field spectrum, whichever is harder. If it is set by the field (as it is in this study), the variation of the high-energy index of the $P(\nu',\theta')$ spectrum in the range $\sim$2.5--3.5 is real and is associated with the varying viewing angle only, neither with the changes of the magnetic field spectrum, nor with the electron energy distribution. The slope of the field spectrum at $k<k_B$ plays a minor role for the low-energy slope of the radiation spectrum. Note that the variation of the lower-energy index between 1 and 0 (or 0 and --1 for the photon spectrum) is a {\em benchmark} feature of the jitter radiation regime. 

Regarding jitter radiation, the reader we shall be warned on the following. Recently an invalid statement has been made \citep{Fleish06} that the spectrum $F_\nu\propto \nu^1$ below the peak, ``valid in the presence of ordered small-scale magnetic field fluctuations, does not occur in the general case of small-scale random magnetic field fluctuations." It has been demonstrated \citep{M05,M06} that the statement is flawed since jitter radiation from {\em random} magnetic fluctuations with a fairly general distribution function (not just ``ordered small-scale'' fields) does allow for $\propto \nu^1$ spectra. Moreover, the entire range from $\propto \nu^1$ to $\propto\nu^0$ is allowed for  monoenergetic electrons, and even softer spectra can form, depending on the electron energy distribution. Another confusing issue is related to the absorption-like $\propto\nu^2$ spectrum \citep{Fleish06}. Such a spectrum is due to plasma dispersion and shall be seen in/near the radio band, not in gamma-rays, as the reader might incorrectly infer \citep{M05}. We stress once again here that the spectral shape of jitter radiation depends, in general, on the viewing angle: it can vary between $\propto \nu^1$ to $\propto\nu^0$, and this is the effect we use in the present study.

\begin{figure}
 \centering
 \plottwo{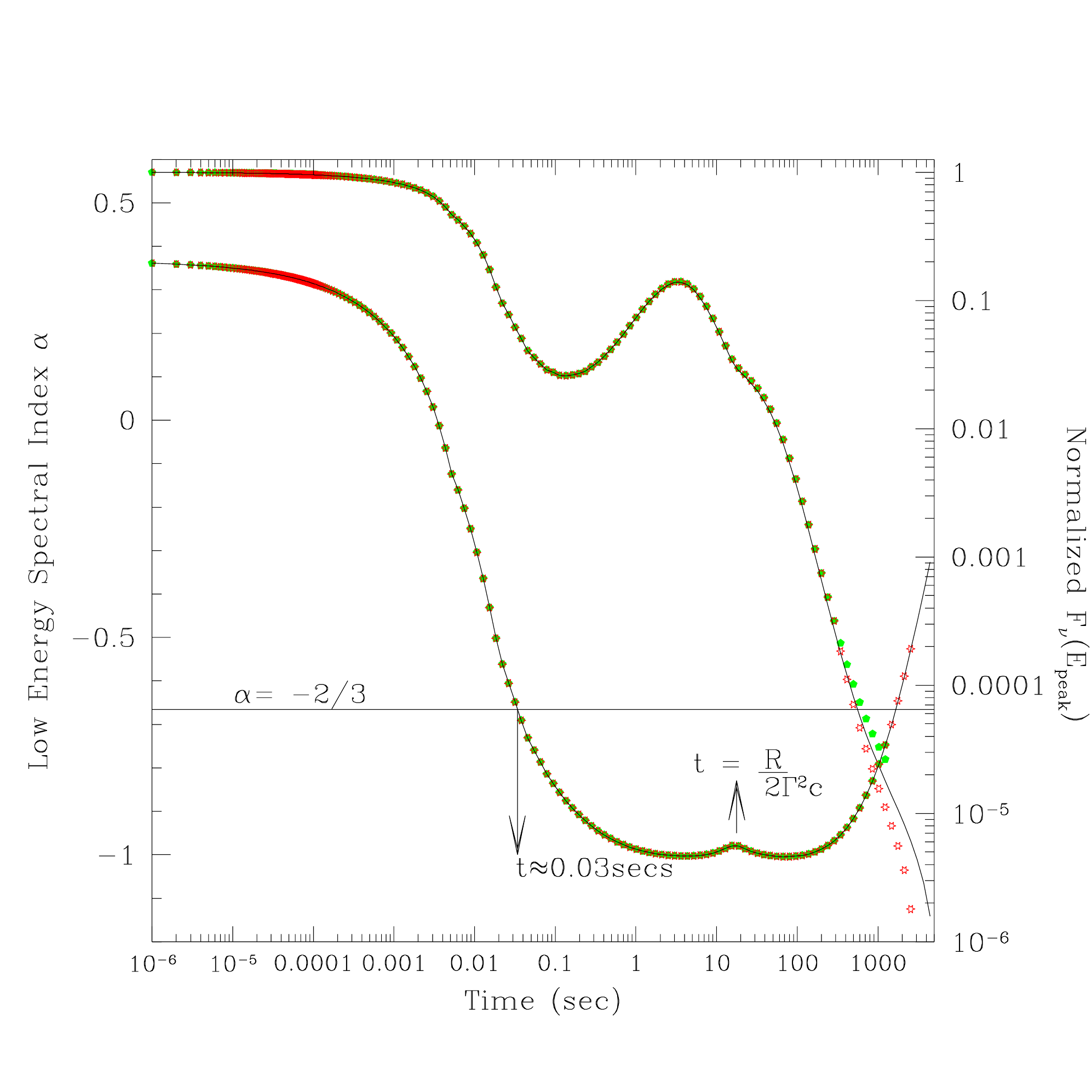}{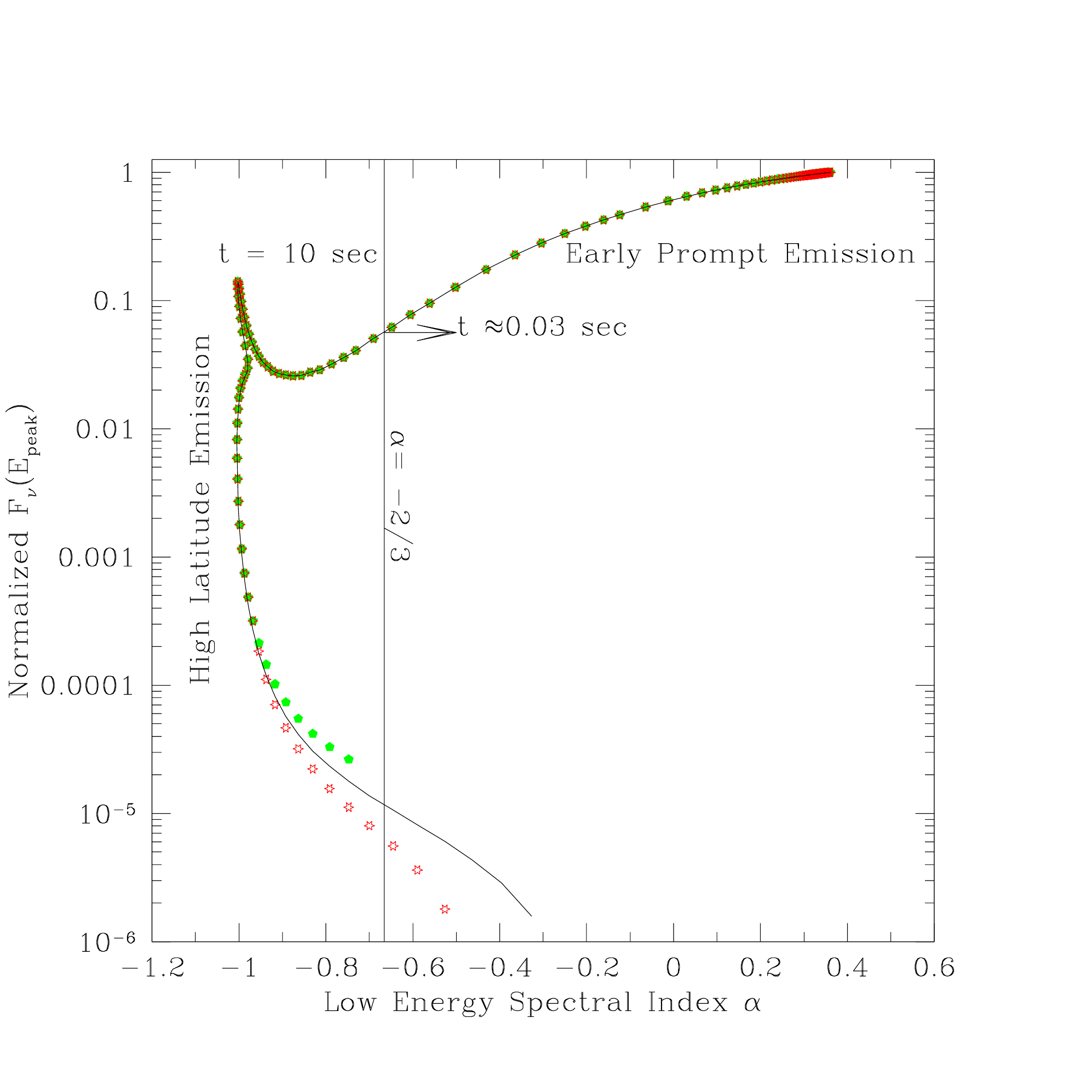}
%\plotone{Single_Spike_fluxavst2}
%\vspace{-1.5cm}
%\plotone{Single_Spike_FEpvsalpha}
 \caption{ {\it (left panel)} --- The single pulse light-curve -- the time evolution of $\alpha$ (bottom curve, left axis) and normalized flux, $F_{\nu}(E_{\rm peak})$ (top curve, right axis); 
 {\it (right panel)} --- the 
correlation of $F_{\nu} (E_{\rm peak)}$ and $\alpha$.  Three cases are plotted: $\theta_{\rm obs}=0$ (green dots), $\theta_{\rm jet}/2$ (red dots), and $\theta_{\rm jet}$ (solid line). The feature in $F_{\nu}(E_{\rm peak})$ at $t\sim R/(2c\Gamma^2)\sim 10$~s is due to jitter spectrum feature at oblique angles $\theta\sim1$; it is a prediction of the model. The overall tracking behavior is evident. The synchrotron-violating spectra are seen at early times $t<0.03$~s and at very late times $t >1000$~s, near a jet break. In the right panel, the upper part of the path may be interpreted as the early emission phase of an individual emission episode and the vertical segment is the high-latitude emission phase. Steepening of the spectrum at very late times is another observational prediction of this model.}
 \label{avst}
  \label{tracking}
\end{figure}

\begin{figure}
 \centering
\plottwo{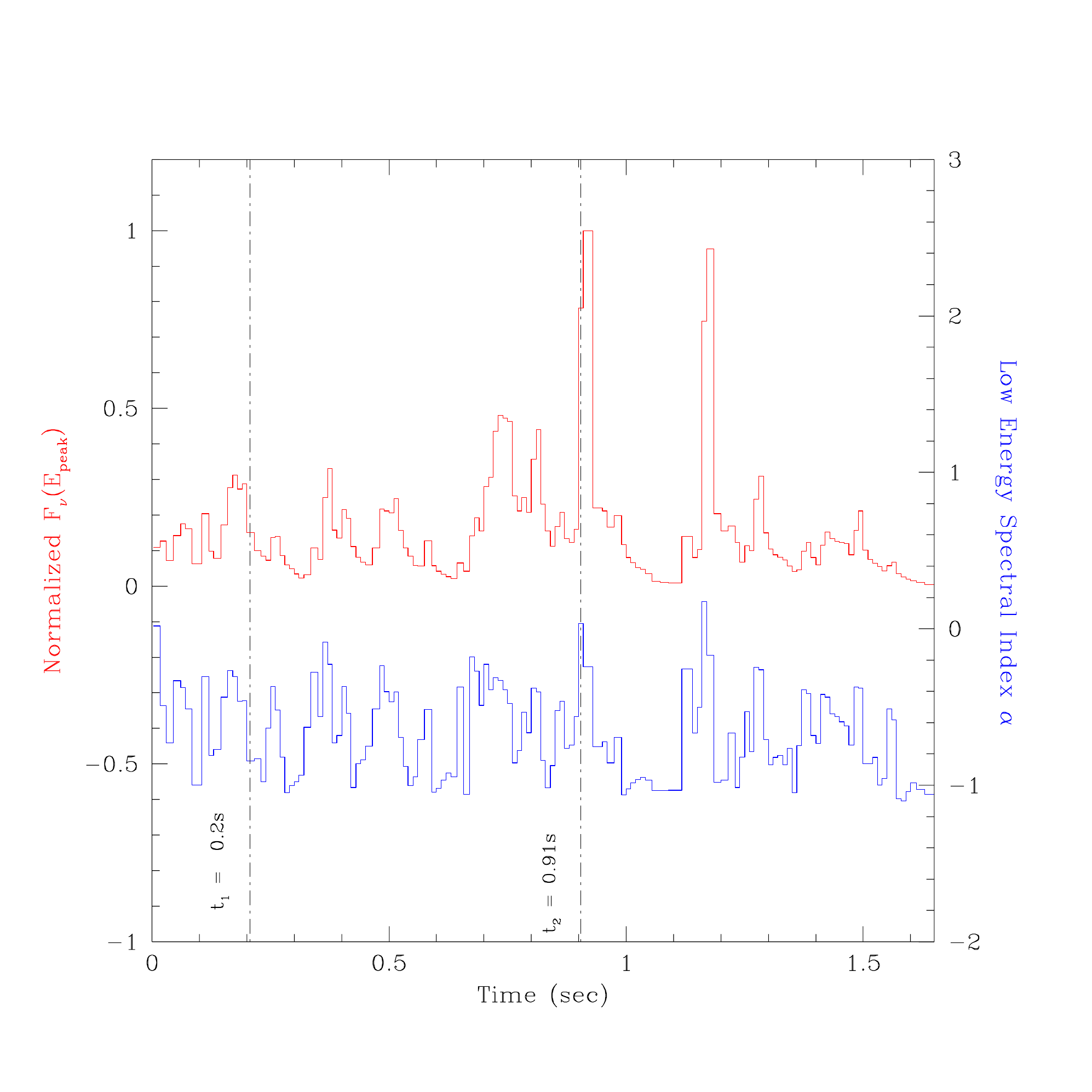}{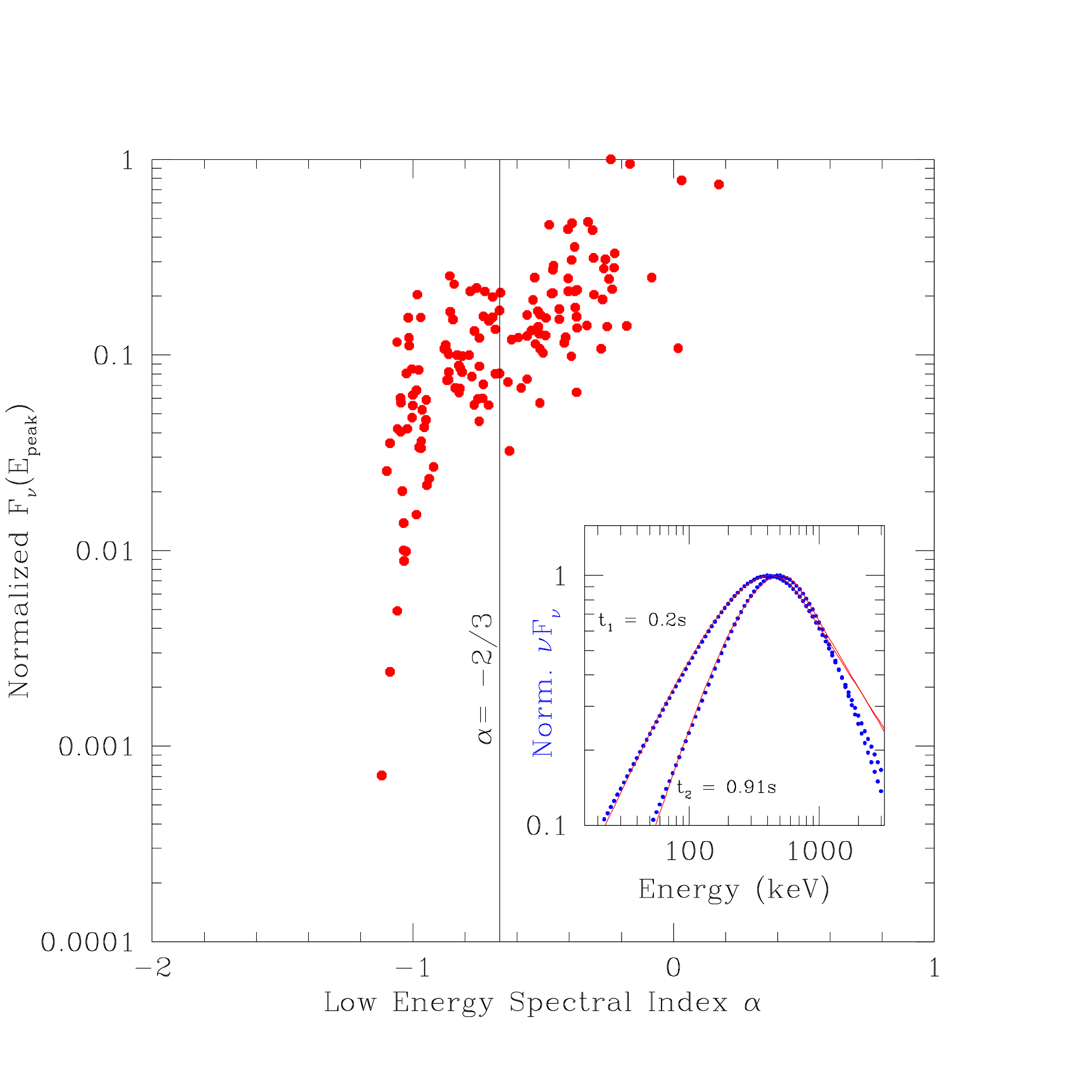}
%\plotone{Sharp_LowBin_FEp_alpha}
%\vspace{-1.5cm}
%\plotone{Scatter_and_Fit}
 \caption{{\it (left panel)} --- The binned synthetic energy flux light-curve with 100 emission episodes (red curve, left axis) and the low-energy index $\alpha$ versus time (blue curve, right axis); 
 {\it (right panel)} --- Scatter plot of $\alpha$ vs. normalized flux $F_{\nu}(E_{\rm peak})$. The tracking evolution and the positive correlation of $F_{\nu}(E_{\rm peak})$ and $\alpha$ are evident. Most of the points are clustered at $\alpha\sim -1$ and there is a smaller population of synchrotron-violating points with $\alpha>-2/3$ --- both results are in excellent agreement with BATSE spectral data \citep{Kaneko+06}. The inset shows two representative spectral fits (red curves) to the model (blue points) at times corresponding to low and high $\alpha$'s; these times are marked by vertical dot-dashed lines in the left panel. The fits are accurate everywhere, except the high energy edge.}
 \label{syntlc}
 \label{syntscatter}
\end{figure}

The obtained light-curve $F_{\nu}(t)$ represents the full spectrum evolution in time for each individual pulse of a prompt GRB and, possibly, an X-ray flare. In order to analyze spectral correlations and to compare our model to observational results, we fit the spectrum at each time step, $t$, with the Band function \citep{Band93} to obtain the peak energy $E_{\rm peak}$ (or, in general, the break energy), the low- and high-energy {\em photon indices} $\alpha$ and $\beta_{\rm spec}$ and the overall normalization $A$. The fit is done in the linear scale of $E$ (not $\log(E)$). This makes the fit more sensitive to the spectral shape around $E_{\rm peak}$ and less sensitive to the shape at energies lesser and greater than $E_{\rm peak}$. This is a better method, since it is known that in the region $E < E_{\rm peak}$, the spectral shape is not a simple power law. Many GRB spectra were analyzed similarly, hence the comparison with observations is facilitated. Since noise is absent in our spectra, in contrast to observations, the $\chi^2$ values should not be compared. In order to mimic the spectral limitations of BATSE, we limit our six-decade spectrum to the two-decade spectral window including most of $E_{\rm peak}$-values, except for very soft spectra at late times. In our correlation analysis, we use the low-energy photon index $\alpha$ and the energy flux at the peak energy, $F_{\nu}(E_{\rm peak})$, because the latter is a well-defined physical quantity characterizing an emission mechanism, in contrast to the flux within a fixed energy band often used in GRB data analysis. A detailed study of spectral correlations between other parameters goes beyond the scope of this paper and will be presented elsewhere (Pothapragada, et al. 2009, in preparation).

\section{Results and discussion}

In our model, $\theta_{\rm jet}$ and $t_c=R/(2c\Gamma^2)$ are free parameters; they were chosen to be $t_c\sim10~{\rm s}$ and $\theta_{\rm jet}=5^\circ$; here we also reset $t$ to be the `time since trigger', i.e., the very first photons arrive at $t=0$. Figure \ref{avst}(left)  shows the time evolution of $\alpha$ and $F_{\nu}(E_{\rm peak})$ within a single pulse (emission episode) for three misalignment angles $\theta_{\rm obs}=0,\ \theta_{\rm jet}/2,\ \theta_{\rm jet}$, the third case appears between the other two because we plot the normalized (not absolute) flux. The differences are minimal: the $\alpha$-curves overlap and the $F_{\nu}(E_{\rm peak})$-curves start to deviate around the jet break at $\sim100 t_c$. The overall tracking behavior is evident: the higher flux in the beginning of the pulse corresponds to the larger $\alpha$, i.e., the steeper spectrum below $E_{\rm peak}$. At early times and at very late times (near the jet break) $\alpha$ exceeds the ``synchrotron line of death" $\alpha=-2/3$. This is a prediction of our model. Figure \ref{tracking}(right) represents correlation of the same spectral parameters. The pulse starts off in the upper right corner and follows the upper part of the curve to the left --- this is the early emission phase. At later times $t\gg t_c$, the path turns downward --- this is the high-latitude emission phase. At very late times, one sees steepening of the spectrum.  Because the magnetic field lies predominantly in the shell plane (in the internal shock scenario), we suggest that the polarization of gamma-ray emission shall increase at times when one sees the shell nearly edge-on in the comoving frame,  i.e., at the time $t\sim t_c=R/(2c\Gamma^2)$. A stronger polarization is expected for a sufficiently misaligned jet, $\theta\sim1/\Gamma < \theta_{\rm jet}$, or a distorted internal shock front/reconnection layer. The presence of a feature in $F_{\nu}(E_{\rm peak})$ at $t\sim t_c$ is another model prediction, though the exact shape can depend on the actual magnetic field distribution. 

Finally, we create a synthetic GRB by overlaying 100 individual pulses (now with $t_c\sim1$~s) at random positions and with random amplitudes drawn from a power-law distribution derived from the pulse statistics \citep{Beloborodov+00}. In order to mimic the time-resolved spectral analysis done by the BATSE team, we then bin it in time so that fluence is the same in all bins (the bin sizes, of course, vary); the binned energy flux light-curve is shown in figure \ref{syntlc}(left) with the red line. The bin-averaged spectra are then fit as usual and the dependence of the soft photon index $\alpha$ on time is overplotted in the same graph. The inset in the right panel shows that the synthetic spectra are well fit with the Band function.  Clearly, $\alpha$ tracks flux in the same way it does in figure \ref{trigger}. The scatter plot, figure \ref{syntscatter}(right), does show a positive $F_{\nu}(E_{\rm peak})$-$\alpha$ correlation with some scatter. One can also see that there is a small population of spectra with steep (even steeper than synchrotron) $\alpha$'s and that many spectra have their soft indices clustered around $\alpha=-1$, in great agreement with observational data \citep{Preece+00,Kaneko+06}. We stress that this jitter model is the only one, to our knowledge, which can explain these facts altogether.

\section{Conclusions} 

We developed a model of prompt GRB spectral variability built upon physical understanding of the internal structure of a dissipation region. The model presented here is rather crude, because it uses very simplified models of the magnetic field spatial spectrum, the electron distribution and the simplified relativistic shell kinematics. Yet, the results obtained mimic the observed spectral correlations of the BATSE catalog remarkably well without any {\em ad hoc} assumptions, thus suggesting robustness of the model. We extend our analysis to other spectral parameters and use more elaborate models in forthcoming publications. The following predictions can be made. First, one shall expect steepening of the spectral index below the peak energy at very late times. Such observations are difficult because an afterglow sets in before this and because the peak shifts to low frequencies, where (self-)absorption may play a role. Second, a feature in the light curve $F_{\nu}(E_{\rm peak},t)$ can be expected around $t\sim R/(2c\Gamma^2)$. Third, if the jet is misaligned with $\theta_{\rm obs}\sim1/\Gamma$ and/or the shock front/reconnection layer is distorted, one can expect enhancement of polarization of gamma-ray emission at times $t\sim R/(2c\Gamma^2)$. In magnetic outflows, the ambient field can also contribute to polarization and (synchrotron) emission.

\acknowledgements

We thank S.K. Graham for valuable help with computational issues and for discussions and the referee for numerous valuable and insightful suggestions.
This work has been supported by NSF grant AST-0708213, NASA ATFP grant NNX-08AL39G, Swift Guest Investigator grant NNX-07AJ50G and DOE grant  DE-FG02-07ER54940.

\end{document}